\documentclass{mn2e}

\usepackage{psfig,graphicx,epsfig}

\newcommand{\ssst}{\scriptscriptstyle}
\newcommand{\E}[1]{$\times 10^{#1}$}

\newcommand{\RA}[3]{\mbox{${#1}^{{\rm h}}{#2}^{{\rm m}}{#3}^{{\rm s}}$}}
\newcommand{\Dec}[3]{\mbox{${#1}^{\circ}{#2}'{#3}''$}}
\newcommand{\witherr}[2]{\mbox{$^{+#2}_{-#1}$}}
\newcommand{\err}[1]{$\pm{#1}$}

        \newcommand{\NH}{$N_{\ssst\rm H}$}
\newcommand{\no}{$n_{\ssst 0}$}

\newcommand{\xray}{X-ray}       
\newcommand{\rosat}{{\sl ROSAT}}

\newcommand{\xmm}{{\sl XMM}-{\sl Newton}}

\newcommand{\snr}{{Cygnus Loop}}

\title[thermal conduction in Cygnus Loop]{The origin of the X-ray emitting plasma in the eastern edge
of the Cygnus Loop}
\author[Zhou et al.]{X.\ Zhou$^{1,2}$
 F.\ Bocchino$^{2}$, M.\ Miceli$^{3,2}$, S.\ Orlando$^{2}$, and Y.\ Chen$^{1,4}$\\
$^{1}$Department of Astronomy, Nanjing University, Nanjing 210093, China\\
$^{2}$INAF-Osservatorio Astronomico di Palermo, Piazza del Parlamento 1, 90134 Palermo, Italy\\
$^{3}$Dip. di Scienze Fisiche \& Astronomiche, Univ.\ di Palermo, Piazza del Parlamento 1, 90134 Palermo, Italy\\
$^{4}$Key Laboratory of Modern Astronomy and Astrophysics (Nan-jing University), Ministry of Education, China
}
\begin{document}

\date{Accepted. Received}

\pagerange{\pageref{firstpage}--\pageref{lastpage}} \pubyear{}

\maketitle

\label{firstpage}

\begin{abstract}
The \snr\ is interacting with a protrusion of the cavity wall in its eastern edge (the XA region), where the X-ray emission is very bright. The complexity of the environment and the non-linear physical processes of the shock-cloud interaction make the origin of the \xray\ emission still not well understood.
  Our purpose is to understand the physical origin of the \xray\ emission in the XA region, addressing, in particular, the role of thermal conduction in the interaction process.
  We analyzed two \xmm\ data sets, performing image analysis and
  spatially resolved spectral analysis on a set of homogeneous regions.
  We applied a recently developed diagnostic tool to compare spectral
  analysis results with predictions of theoretical models, and to
  estimate the efficiency of thermal conduction on the X-ray emitting
  shocked plasma.
  We found that the inhomogeneous cavity wall contains both large clumps (the protrusion) and small isolated clumps with different densities. A large indentation bent over to the south is detected. The abundance of the surrounding ISM is $\sim0.2$ times solar value. We confirmed the important role of thermal conduction in the evolution of \xray\ emitting plasma during shock-cloud interaction.
\end{abstract}


\section{Introduction}

The \snr\ is considered to be a proto-typical evolved supernova remnant (SNR). Located at a distance of 540\witherr{80}{100} pc (Blair \& Sankrit 2005), the remnant can be studied with high spatial resolution.
Its surface brightness is also high, which enables us to attain high signal-to-noise ratio observations.
The \snr\ is therefore a great laboratory to study the evolution of a SNR at different scales.

The X-ray emission of the \snr\ is limb-brightened and clumpy, which is
thought to be the result of a recent interaction with an inhomogeneous
cavity wall (Levenson et al.\ 1997, Levenson et al.\ 1999). Several X-ray
bright features in the edge of the remnant were studied, in the framework of the scenario of shock-cloud interaction (Miyata \& Tsunemi 2001,
Levenson et al.\ 2002, Levenson \& Graham 2005, and Patnaude et al.\
2002). However, the issue of the origin of the hot plasma contributing
to the \xray\ emission in these regions was not addressed in detail yet.
One of the difficulty is to take into proper account typical non-linear
physical processes occurring in these systems, such as radiative losses
and thermal conduction.

To better understand the origin of the \xray\ emission, a detailed model that could be accurately compared with the observation is required. Miceli et al.\ (2006) studied an isolated \xray\ knot in the Vela SNR (Vela Fi1D) by directly comparing models with observations. They built up a set of hydrodynamic simulations of shock-cloud interaction including thermal conduction and radiative cooling effects. The model setups were determined taking account of the earlier analysis of the X-ray data (Miceli et al.\ 2005). They synthesized the X-ray emission (maps and spectra) from the models, and compared it directly with the observation for the first time. In this way, a self consistent picture of the hot X-ray emitting plasma was derived, and the importance of thermal conduction was confirmed in this case.
\begin{figure*}
\centerline{ {\hfil\hfil
\psfig{figure=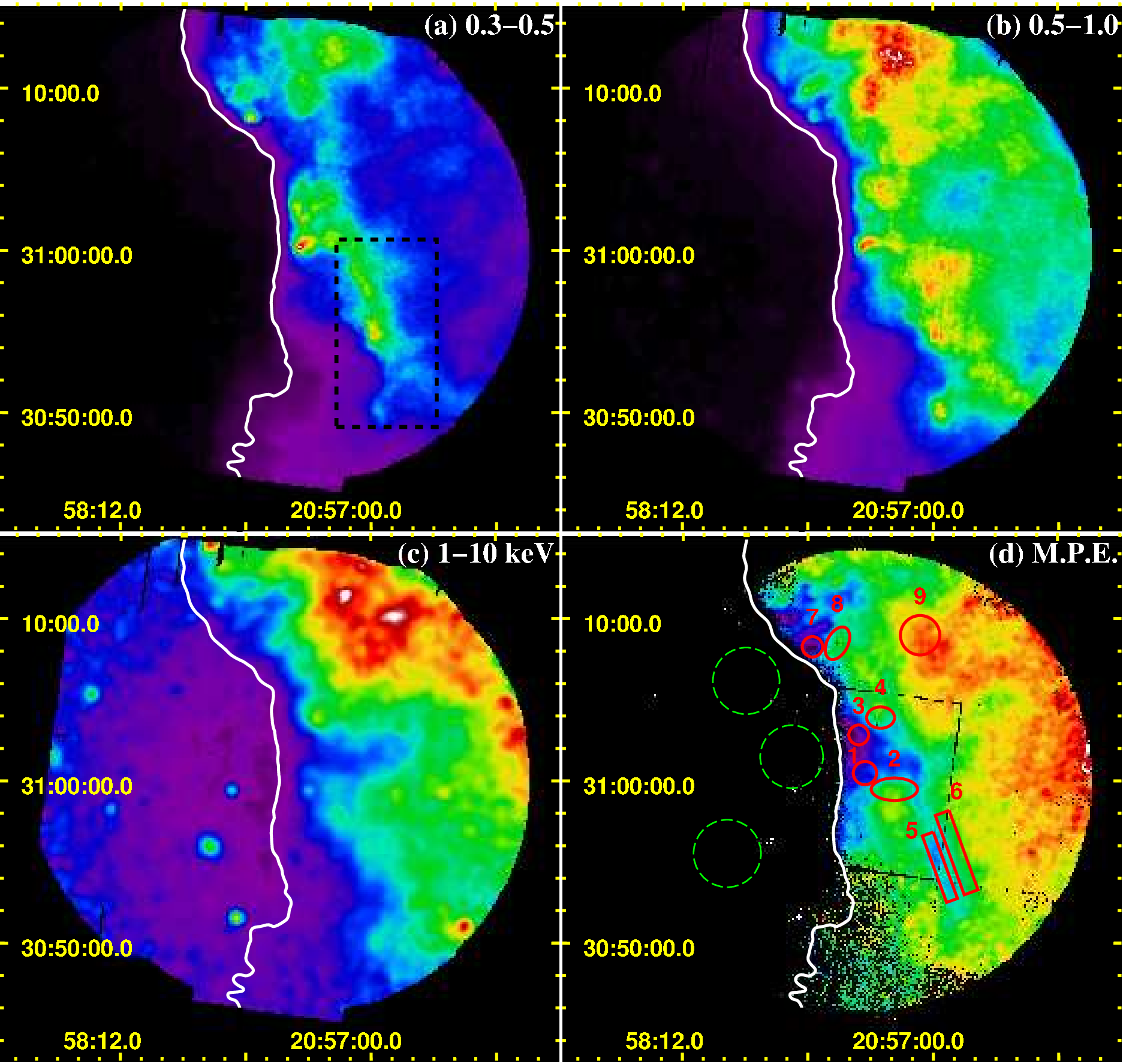,height=14cm,angle=0, clip=} \hfil\hfil}}
\caption{EPIC count rate maps (MOS-equivalent) in three energy bands and median photon energy map. The count rate maps were produced by weighted averaging the MOS1, MOS2, and pn images in the soft (0.3--0.5 keV, {\sl upper-left panel}), medium (0.5--1 keV, {\sl upper-right panel}), and hard (1--12 keV, {\sl lower-left panel}) bands. The maps were adaptively smoothed to a signal-to-noise ratio of 25. The peak values in the three maps are: 1.7\E{-3} ${\rm cnt}~{\rm s}^{-1}~{\rm bin}^{-1}$, 1.6\E{-3} ${\rm cnt}~{\rm s}^{-1}~{\rm bin}^{-1}$, and 2.1\E{-4} ${\rm cnt}~{\rm s}^{-1}~{\rm bin}^{-1}$, respectively.
The dashed rectangular region in the soft band map indicates the extension of the strip (see text).
The median photon energy map ({\sl lower-right} panel) was produced from a combination of the MOS1 and MOS2 data in 0.3--2 keV band. The image was smoothed with a gaussian distribution with $\sigma=8.5\arcsec$, and the minimum number of counts per pixel is 4. The regions chosen for spectra extraction are indicated in red, while the dashed green circles indicate the background regions.
The bin size of each image is 4\arcsec. The white contour marks the $3\sigma$ confidence level of the \xray\ emission in 0.3--2 keV band.
} \label{f:3bandsmpe}
\end{figure*}

Orlando et al.\ (2005) and Orlando et al.\ (2006) examined the shock-cloud
interaction in different physical regimes by using a detailed hydrodynamic
model, including thermal conduction and radiative losses, and synthesizing
the relevant \xray\ emission. Based on these works, Orlando et al.\ (2010) proposed
a diagnostic tool to derive quickly the efficiency of thermal conduction
from the results of the \xray\ data analysis, without the need of running a time consuming ad-hoc hydrodynamical model.  By using such advanced
method, the origin of X-rays in the \snr\ can be investigated to derive a
better understanding of the shock-cloud interaction.  In this paper we
apply this method to a still not well understood region of the \snr,
an outstanding interaction region named ``XA" by Hester \&
Cox 1986. Both bright \xray\ emission and bright optical emission are
present in this region.  The scenario of the blast wave interacting
with a protrusion from the cavity wall was proposed to explain the
properties of optical emission (Danforth et al.\ 2001, Hester \& Cox
1986). The proposed scenario is entirely different from the case of interaction with an isolated
cloud in the Vela SNR studied by Miceli et al. (2005, 2006): therefore, it is necessary
to see whether thermal conduction also plays an important role in this
situation. We analyzed the \xmm\ European photon imaging camera (EPIC)
observations of the XA region to examine
the physical processes that account for the observed \xray\ emission.  The observations and results are described in \S2 and \S3. In
\S4, we discuss the interpretation and derive the characteristic parameters
of shock-cloud interaction. The conclusions are summarized in \S5.

\section{Observation and Data Reduction}

\begin{table}{}
\caption{Observational parameters\label{tb:obs}}
\medskip
\centering
\renewcommand{\footnoterule}{}
\begin{minipage}{8.0cm}
\begin{tabular}{lccc} \hline
 & MOS1 & MOS2 & pn  \\ \hline
$t_{exp}$\footnote{exposure time after screening of observations 0018141101/0018141301.} (ks) & 7.7/10.7 & 7.5/10.7 & 4.1/4.4 \\
filter  & medium & medium & thin1 \\

\hline
\end{tabular}
\end{minipage}
\end{table}

We here analyze \xmm\ archive observations (IDs: 0018141101 and 0018141301, PI: T.\ Gaetz). These two observations were performed with the EPIC MOS and EPIC pn cameras in full frame mode, pointing at ($\RA{20}{57}{21}.17$, $\Dec{31}{00}{10}.5$) (J2000). They were performed on 2 November 2002 and 12 November 2002, respectively. The EPIC cameras offer a spectral resolution $E/\Delta E \sim 20$--50 in the energy range from 0.15 to 15 keV. The field of view (FoV) is $\sim30\arcmin$,  and the angular resolution is $6\arcsec$ FWHM. The main information about the data are listed in Table~\ref{tb:obs}.

All data were processed using the Science Analysis System (SAS) V9.0.0. The original events files were screened following the method suggested by Snowden \& Kuntz (2007), and we performed the correction of vignetting effects by using the task {\sl EVIGWEIGHT} (Arnaud et al.\ 2001). The PATTERN and FLAG values were set to zero.

\section{Results}

\subsection{X-ray morphology}

\begin{figure}
\centerline{ {\hfil\hfil \psfig{figure=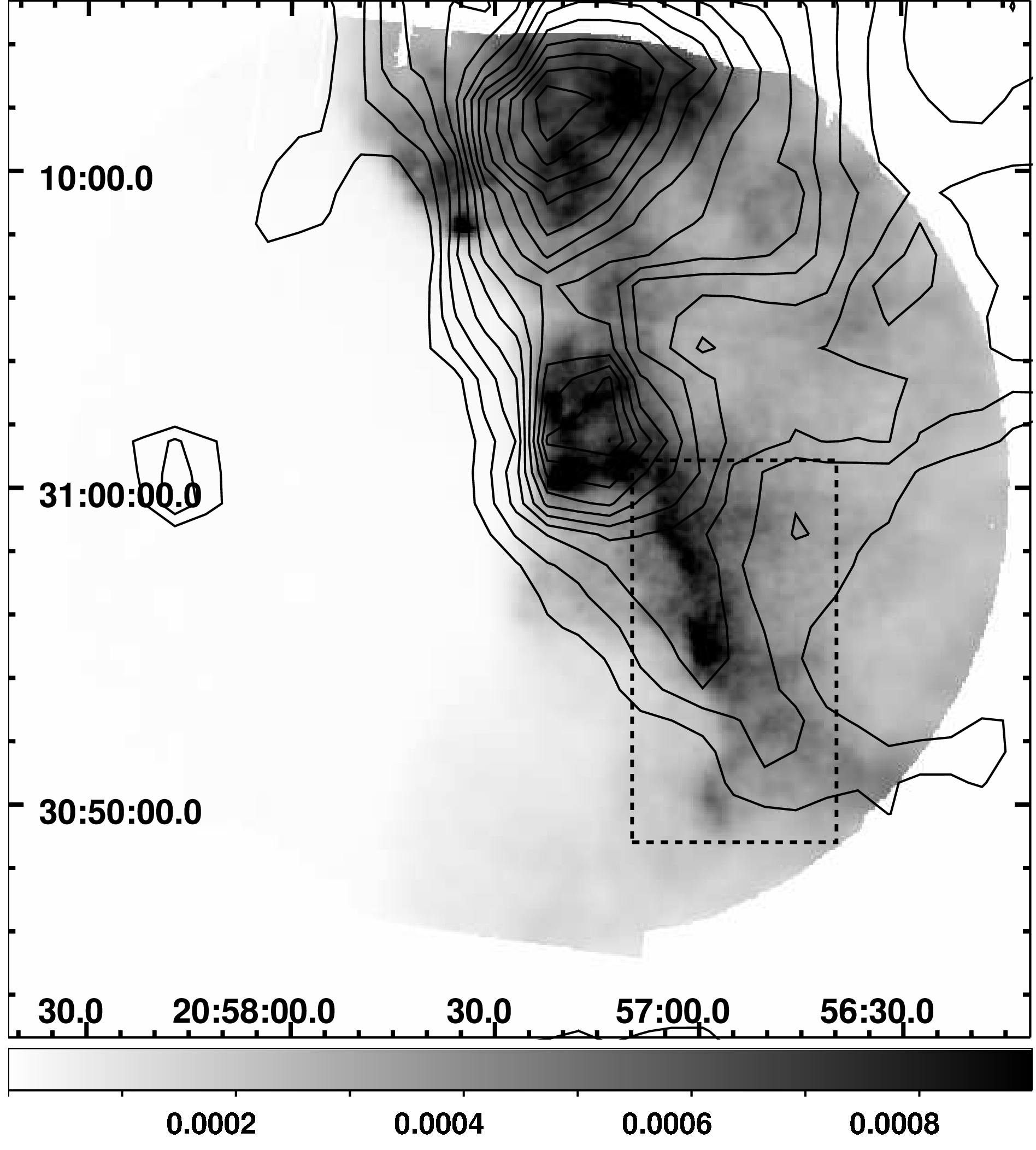,height=8.0cm,angle=0,clip=} \hfil\hfil}}
\caption{\xray\ count rate map in the 0.3--0.5 keV band (the same as in Fig.~\ref{f:3bandsmpe}a), with 25 $\mu$m contours derived from IRIS data superimposed. The contour levels are from 2.125 MJy $sr^{-1}$ to 2.125 MJy $sr^{-1}$ in step of 0.053 MJy $sr^{-1}$. The effective resolution of the IRIS 25 $\mu$m data is $\sim 3{\arcmin}$.8. The dashed rectangular region is the same as in Fig.~\ref{f:3bandsmpe}a.} \label{f:softvsIR}
\end{figure}

The count rate maps in the 0.3--0.5 keV, 0.5--1 keV, and 1--12 keV bands
shown in Fig.~\ref{f:3bandsmpe} are weighted averages of
the MOS1, MOS2, and pn images (MOS-equivalent, as in Miceli et al.\
2005). In the western side the emission is confined by a clear
curved border (indicated by the white contour in Fig.~\ref{f:3bandsmpe}), well visible in all the three bands, which could be considered as the edge of the
remnant. The soft band image is mainly characterized by several clumpy
features and a bright long strip extending from the center of the FoV to
the southwest. Fig.~\ref{f:softvsIR} shows the {\sl Infrared Astronomical Satellite (IRAS)} contours at 25 $\mu$m wavelength from Improved Reprocessing of the {\sl IRAS} Survey (IRIS; Miville-Deschenes \& Lagache 2005) superimposed onto the \xray\ image in the 0.3--0.5 keV band. An
enhancement of the mid-IR emission is well correlated with the \xray\
strip (see dotted box in Fig.~\ref{f:softvsIR}). In the \xray\ medium
band (Fig.~\ref{f:3bandsmpe}b), the emission is more
extended, and mainly distributed behind the soft band clumps, while the hard
band emission (Fig.~\ref{f:3bandsmpe}c) is more diffuse and farther from the border of the shell. A
large bright region is visible to the northwest in all of the three band
images.

A high resolution median photon energy ($\overline{E}$) map was produced from the MOS1 and MOS2 event files in the 0.3--2 keV band (Fig.~\ref{f:3bandsmpe}d). The bin size is $4\arcsec$, and the minimum number of counts per pixel is 4. The map was smoothed with a gaussian distribution with $\sigma=8.5\arcsec$. A noteworthy structure is the central low $\overline{E}$ region corresponding to the ``dense finger" protruding from the cavity wall observed in the optical bands (Danforth et al.\ 2001). Moreover, a strip protruding from the tip of the ``dense finger" to the southwest is found, which corresponds to the \xray\ strip in the 0.3--0.5 keV band (see Fig.~\ref{f:softvsIR}).

Four couples of regions were chosen for spectra extraction; each couple includes a soft band bright knot and the diffuse region (in the medium band) immediately behind it (regions 1-2, 3-4, 5-6, and 7-8 in Fig.~\ref{f:3bandsmpe}d). In addition, a region to the northwest
was also selected (region 9). 
The median photon energy map is a
good indicator of the spectral distribution (Hong et
al.\ 2004). Therefore, the regions were also selected as
homogeneous as possible in the $\overline{E}$ map to avoid large
spread of spectral properties.  The spectra from the regions surrounding
shocked clouds are sensitive to the physical state of the interaction
(Miceli et al.\ 2006) and Orlando et al.\ (2010) found that the spectral
analysis of regions selected in the this way is an efficient diagnostic tool to
investigate the role of thermal conduction (see Sect. \ref{The role of thermal conduction}).

\begin{figure}
\centerline{ {\hfil\hfil \psfig{figure=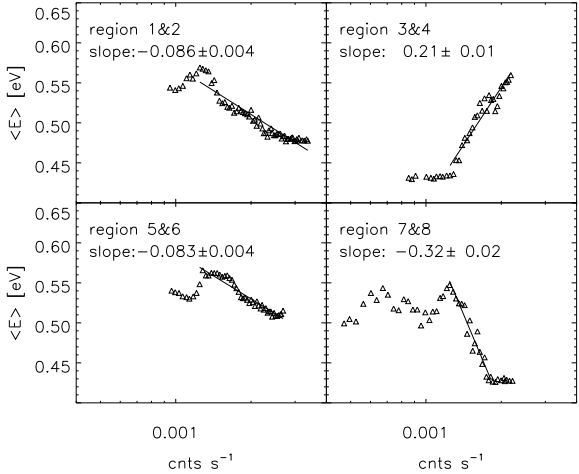,height=7.0cm,angle=0,clip=} \hfil\hfil}}
\caption{Median photon energy ($\overline{E}$) versus count rate ($R$) plot for the regions defined in Fig.~\ref{f:3bandsmpe}d.
Linear fittings were performed on the point sets above 1.2\E{-3} cnts s$^{-1}$ (also below 1.8\E{-3} cnts s$^{-1}$ for region 7\&8). The corresponding best-fit slopes are shown in each figure.}\label{f:RvsE}
\end{figure}

The relationship between the median photon energy and the count rate ($R$) is critically influenced by different physical processes such as thermal conduction in the shock-cloud interaction and can provide important information (Orlando et al.\ 2010). In this framework, we derived the $\overline{E}$ versus $R$ scatter plots for each of the four couples of regions (Fig.~\ref{f:RvsE}). To produce these plots, the range of $R$ was divided into 100 channels equispaced on linear scale. Then, we derived the $\overline{E}$ by averaging the median photon energy in all the pixels (in the selected regions) with count rate inside a given channel.
The points above $\sim$ 1.2\E{-3} ${\rm cnt}~{\rm s}^{-1}~{\rm bin}^{-1}$ for all the couples of regions, except regions 3\&4, show a uniform descending trend. The descending slopes are very flat in regions 1\&2 and 5\&6, and steeper in region 7\&8 where a plateau is also present above $\sim$ 1.8\E{-3} ${\rm cnt}~{\rm s}^{-1}~{\rm bin}^{-1}$. The ascending trend of $\overline{E}$ vs $R$ in regions 3\&4 suggests that the nature of the \xray\ emission in these two regions is somehow different from the other regions.

\subsection{Spectral analysis}
\label{Spectral analysis}

The \xmm\ data enabled us to perform the spectral analysis with good spatial resolution. The spectra were extracted from the regions defined in \S{3.1}. The MOS and pn spectra of the two observations were fitted simultaneously, and a free parameter was multiplied to each spectra to account for the residual discrepancy of extraction area for different instruments, which was tied for the two observations. We further applied a gain fit to the pn spectra in some regions, in which the residuals between the data and the best fit models fluctuate notably around the strong emission lines.

\begin{table*}
\caption{Spectral-fit parameters. The \NH\ parameters are fixed to 9\E{20} cm$^{-2}$. The errors are at $90\%$ confidence levels.\label{tb:fitpara}}
\medskip
\centering
\renewcommand{\footnoterule}{}
\begin{minipage}{17.0cm}
\begin{tabular}{lccccccc} \hline

  & \multicolumn{3}{c}{CIE component} & \multicolumn{3}{c}{NEI component} & \\
 Region & $k$T$_{1}$  &  C, N, O & n$^2_{1}$L$_1$ \footnote{given by $4\pi {\rm EM}/\Theta$, where EM is the emission measure and $\Theta$ is the solid angle subtended by each region.} & $k$T$_{2}$  & $\tau$ & n$^2_{2}$L$_2$ $^a$ & $\chi^{2}$/d.o.f. \\
  &  (keV) & (Z$_{\sun}$) &  ($10^{20}$ cm$^{-5}$) & (keV) & ($10^{12}$ cm$^{-3}$ s) & ($10^{18}$ cm$^{-5}$) &  \\ \hline

1 & 0.090\witherr{0.002}{0.003} & 0.21\witherr{0.02}{0.03} & 14\err{2} & 0.26\witherr{0.02}{0.03} & 0.4\witherr{0.2}{0.3} & 8\err{2} &343.7/290 \\
2 & 0.103\witherr{0.003}{0.004} & 0.14\err{0.02} & 5.2\err{0.5} & 0.26\witherr{0.01}{0.03} & 0.7\err{0.4} & 12\witherr{3}{2} & 389.9/375 \\
3 & 0.074\witherr{0.008}{0.007} & 0.17\err{0.06} & 40\err{10} & 0.26\witherr{0.02}{0.03} & 0.13\witherr{0.06}{0.1} & 5\witherr{1}{3} & 186.1/186 \\
4\footnote{the conjunct abundances of C, N, and O of NEI component are free here, and the best-fit values of regions 4 and 9 are 2.2\witherr{0.5}{0.2} Z$_{\sun}$ and 1.9\witherr{0.2}{0.3} Z$_{\sun}$, respectively.} & 0.114\witherr{0.001}{0.002} & 0.16\err{0.02} & 6.7\err{0.6} & 0.40\witherr{0.02}{0.04} & 0.29\witherr{0.07}{0.1} & 5.5\err{0.9} &349.7/327 \\
5 & 0.102\witherr{0.002}{0.003} & 0.22\witherr{0.02}{0.03} & 6.8\witherr{1}{0.5} & 0.251\witherr{0.008}{0.02} & $>0.5$ & 13\err{3} & 409.5/336 \\
6 & 0.115\err{0.002} & 0.224\witherr{0.008}{0.01} & 3.8\err{0.2} & 0.32\err{0.01} & 0.25\witherr{0.03}{0.02} & 9\err{1} & 504.6/393 \\
7 & 0.084\err{0.005} & 0.24\witherr{0.05}{0.08} & 16\err{5} & 0.24\err{0.02} & $>0.1$ & 6\err{1} & 213.5/194 \\
8 & 0.096\witherr{0.009}{0.01} & 0.26\witherr{0.09}{0.07} & 6\witherr{2}{3} & 0.241\witherr{0.004}{0.007} & $>1.0$ & 24\err{3} & 351.2/283 \\
9$^b$ & 0.151\witherr{0.003}{0.004} & 0.187\witherr{0.01}{0.02} & 2.38\witherr{0.2}{0.07} & 0.46\err{0.02} & 0.21\witherr{0.03}{0.02} & 8\err{1} & 517.5/477 \\
\hline
\end{tabular}
\end{minipage}
\end{table*}

\begin{figure}
\centerline{ {\hfil\hfil \psfig{figure=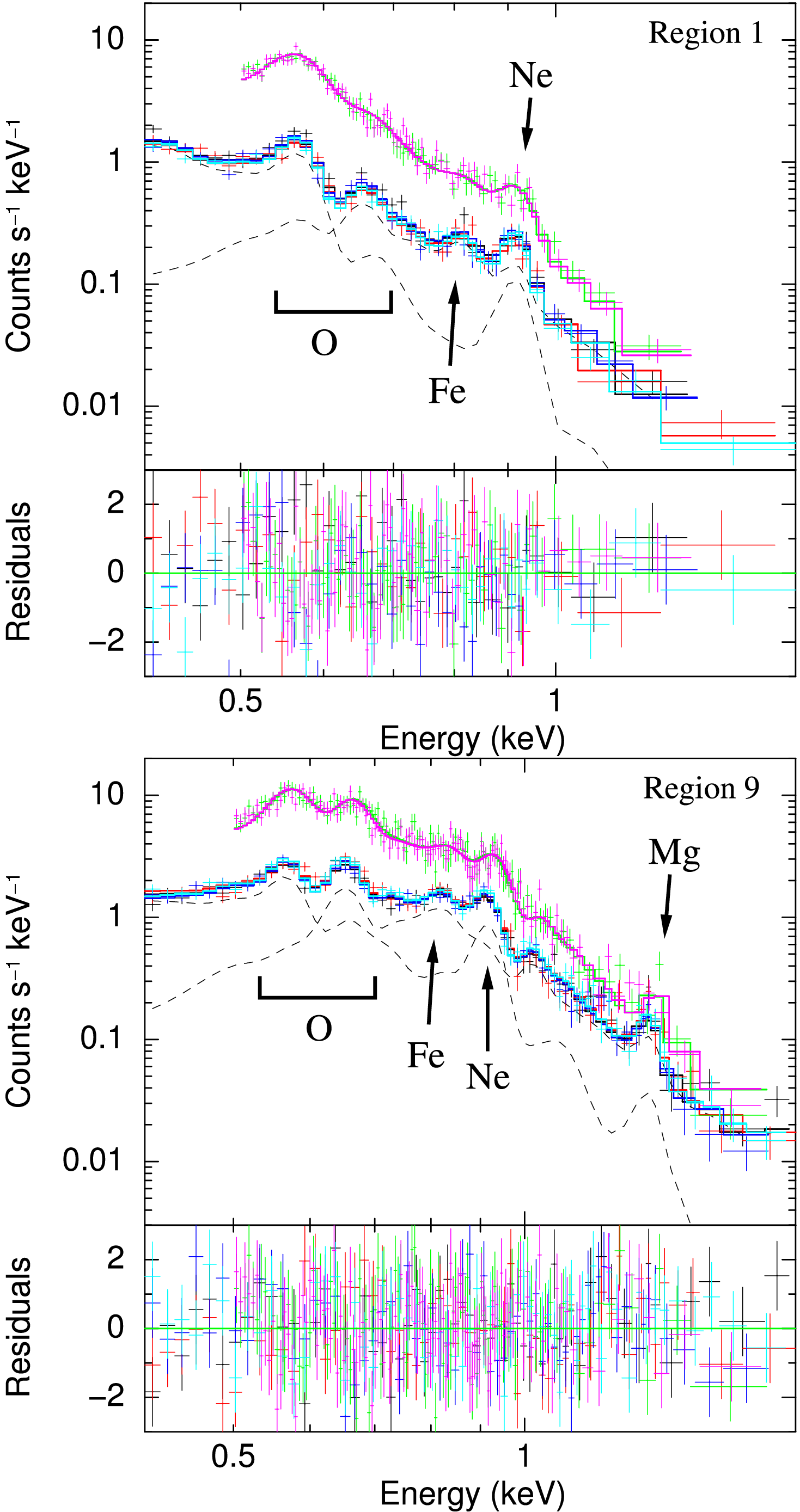,height=14.0cm,angle=0,clip=} \hfil\hfil}}
\caption{MOS ({\sl lower}) and pn ({\sl upper}) spectra extracted from region 1 ({\sl upper panel}) and region 9 ({\sl lower panel}) with the corresponding best fit model and residuals. The individual components are shown in dashed black lines for MOS1 spectrum of observation 0018141101 only. The visible emission lines are labeled (O: 0.51--0.7 keV, Fe: $\sim$ 0.8 keV, Ne: $\sim$ 0.9 keV, and Mg: $\sim$ 1.4 keV).}
\label{f:spec}
\end{figure}

We grouped the spectra to have at least 25 counts per bin, then we fitted
them by using XSPEC V12.5.1. To model the spectra, two thermal components
are required: a hotter component in non-equilibrium of ionization (NEI)
and a cooler component in collisional ionization equilibrium (CIE). The
models VNEI (Borkowski et al.\ 2001) and VMEKAL (Mewe et al.\ 1985, 1986;
Liedahl et al.\ 1995), modified by the interstellar absorption model
WABS (Morrison \& McCammon 1983), were applied here for all spectra. We
verified that other combinations with less free parameters (VMEKAL only,
VNEI only, or two VMEKAL models) do not give a satisfactory description
of the spectra.  At first, we set the \NH\ free, but in some regions its
best fit values are consistent with zero and with very large error bars,
but the values in the other regions are consistent with 9\E{20} cm$^{-2}$
which is typical in this part of the \snr\ (Miyata \& Tsunemi 2001). So
we fixed the \NH\ to 9\E{20} cm$^{-2}$ in all the regions. We set the
abundances of C and N equal to that of O, and left these abundances free
to vary for the cold component.  As for the hot component, the conjunct
abundances of C, N, and O, were found to be consistent with the solar
values in all regions except regions 4 and 9, so we fixed them to the
solar values. However, in regions 4 and 9, we left the CNO abundance of
the hot component free to properly model the spectra.  The fittings were
not improved significantly (according to the {\sl F}-test) by adding an
extra component to the model or leaving more parameters free.
 The fitting results are summarized in Table~\ref{tb:fitpara}. The
 representative spectra extracted from region 1 and 9 are shown in
 Fig.~\ref{f:spec}.

As shown in Table~\ref{tb:fitpara}, the CNO abundances of the cooler component are considerably lower than solar values. A similar result was found in other regions near the edge of the \snr\  (Leahy 2004, Levenson \& Graham 2005, Katsuda et al.\ 2008a, Miyata et al.\ 2007, and Tsunemi et al.\ 2008). The origin of the subsolar values of the abundances around the remnant is still controversial, and it may be related either to metal depletion on to ISM dust grains (Vancura et al.\ 1994) or to local peculiar abundances. 
We found that the CNO abundance of the hotter component in regions 4 and 9 is significantly larger than solar, which is consistent with an ejecta origin. This confirms the different nature of the \xray\ emission in these regions, as expected from the peculiar $\overline{E}$ vs $R$ trend shown in the previous section. Considering the presence of other emission lines, we also tried to set the Ne, Mg, and Fe abundances free, but the fittings were barely improved.

The temperature of the CIE component is the highest in region 9, while it is almost uniform in the other regions. The temperatures of NEI component are almost the same in different regions, except for the regions 4 and 9 where the temperatures are higher than the others.

\section{Discussion}
\subsection{Global properties of the \xray\ emitting plasma}
The patchy morphology of the soft band X-ray emission (see Fig.~\ref{f:3bandsmpe}a) and the filaments observed in the optical band (Danforth et al.\ 2001) strongly indicate that in the XA region the blast wave of the \snr\ is encountering a large scale inhomogeneity of the ISM. In such interacting region, the \xray\ emitting plasma can be typically described with two thermal components associated with two phases of ISM swept up by the shock, as shown in theoretical models (Miceli et al.\ 2006, Orlando et al.\ 2006, 2010). In this scenario, the two temperature components found in our spectral analysis probably indicate two thermal phases of ISM: i) the shocked dense cloud material and ii) the surrounding corona. Analogous features were also reported in the Vela SNR observed with \rosat\ (Bocchino et al.\ 1999) and \xmm\ (Miceli et al.\ 2005), and in several regions of the \snr\ (Levenson et al.\ 2002, Katsuda et al.\ 2008a, Miyata et al.\ 2007, and Uchida et al.\ 2008).

\begin{table}{}
\caption{Derived parameters for each region.\label{tbl:dt}}
\medskip
\centering
\renewcommand{\footnoterule}{}
\begin{minipage}{8.0cm}
\begin{tabular}{lcccc} \hline

region & f$_1$ & $n_1 ({\rm cm}^{-3})$ & $n_2 ({\rm cm}^{-3})$ & $t_{int}$ (kyr)\footnote{the elapsed time of interaction, derived by $\tau/n_2$ where $\tau$ is the ionization time scale of the hot component.}\\ \hline
1 & 0.95\witherr{0.01}{0.02} & 6.4\err{0.7} & 2.20\err{0.02} & 5\witherr{3}{4}\\
2 & 0.87\err{0.03} & 3.9\err{0.4} & 1.54\err{0.03} & 14\err{8}\\
3 & 0.985\witherr{0.006}{0.01} & 11\err{2} & 3.048\witherr{0.006}{0.01} & 1.3\witherr{0.6}{1}\\
4 & 0.91\err{0.02} & 4.4\witherr{0.4}{0.5} & 1.26\witherr{0.01}{0.02} & 7\witherr{2}{3}\\
5 & 0.90\err{0.03} & 4.2\witherr{0.4}{0.4} & 1.70\err{0.03} & $>9$\\
6 & 0.85\err{0.02} & 3.1\witherr{0.2}{0.3} & 1.13\err{0.02} & 7\witherr{0.8}{0.6}\\
7 & 0.97\err{0.01} & 8\err1 & 2.687\witherr{0.009}{0.01} & $>1$\\
8 & 0.80\witherr{0.06}{0.09} & 4.9\witherr{0.9}{1} & 1.94\witherr{0.05}{0.06} & $>16$\\
9 & 0.76\err{0.03} & 2.7\witherr{0.2}{0.3} & 0.90\err{0.03} & 7.4\witherr{1}{0.7}\\

\hline
\end{tabular}
\end{minipage}
\end{table}

The extension of the line of sight (L) through the nearly spherical remnant (with a radius of $\sim84'.5$) at the location of each region could be approximated as the length of the chord there. Then the electron density of the $i$-$th$ component can be derived considering $L_i=L\times f_i$, where $f_i$ is the filling factor. Under the assumption of pressure equilibrium between the two components, the filling factor can be estimated as $f_1=\left(\frac{{\rm EM}_2}{{\rm EM}_1}\frac{{\rm T}_2^2}{{\rm T}_1^2}+1\right)^{-1} $ and $f_2=1-f_1$ (see Bocchino et al.\ 1999). From the ionization time scale and the density of the hot component, we also derived the time elapsed after the interaction with the shock, $t_{int}$.
The results of filling factors, densities, and $t_{int}$ (with corresponding uncertainties) are listed in Table~\ref{tbl:dt}.

Since the filling factors of the low temperature component are predominant in all the regions, the dense cloud material always fills most of the volume. The filling factors and the densities of the cold components are always higher in the knot regions than in the corresponding diffuse regions, and this confirms the inhomogeneities of the ISM and indicates that the knot regions mostly subtend the shocked cloud material.

The $t_{int}$ is longer in the diffuse regions than in the corresponding knot regions, and this is consistent with the direction of propagation of the blast wave, though uncertainties are large. In regions 5, 7, and 8, the hot component seems close to ionization equilibrium, and $t_{int}$ would be maybe too long considering the \snr\  age. The discrepancy may be caused by mistaking the extension along the LoS. As an example, in region 8, if we assume the extension along the LoS to be equal to the size of the region in the plane of the sky, we obtain more reasonable values ($t_{int}>3000$ yr), while the densities become larger (n$_1$=27\witherr{5}{7}, n$_2$=10.7\err{0.3}). So the shocked cloud in regions 7 and 8 is probably a small, compact isolated knot which is much denser than the others. The case could be similar in region 5.

As explained in Sect. \ref{Spectral analysis}, the low temperature component can be associated with the shocked ISM, because of its characteristic low abundances.
The densities of the \xray\ emitting shocked clouds (i.~e. $n_1$ in Table~\ref{tbl:dt}) are comparable with that obtained in other cloud-shock interaction regions in the \snr; for example, in the northeastern limb $n=$1.6--4 cm$^{-3}$ (Miyata et al.\ 2007, the densities of the high temperature component are 1--2 cm$^{-3}$ there), in the southwestern region $n=$0.5--1 cm$^{-3}$ (Patnaude et al.\ 2002), and in the western limb the pre-shock cloud density is 5 cm$^{-3}$ (Levenson et al.\ 2002). The densities of FilD, a shocked cloud in the Vela SNR, were found to be 0.32--5 cm$^{-3}$ for the cloud core and 0.14--5 cm$^{-3}$ for the corona (Miceli et al.\ 2005).

The overabundances revealed in the hot components in region 4 and region 9 are consistent with an ejecta origin, and their temperatures are higher than that of the shocked ISM derived in the other regions. Both region 4 and region 9 are bright in the hard band emission. In region 4, a bright knot is visible in the hard band map (Fig.~\ref{f:3bandsmpe}c), and this may indicate a fragment or protrusion in the ejecta. In region 9, the temperature of the cooler component is higher than that in the other regions, and this is consistent with its lower density. The region is located inside the remnant, where the ejecta is likely to be prominent ($<$~0.85~R$_{\rm shock}$, Tsunemi et al.\ 2007, Katsuda et al.\ 2008b). Indeed, in the large \xray\ bright region to the northwest of our FoV (containing region 9) there is also the contribution of some edge-on sheet of shocked ISM (Raymond et al.\ 1988, Blair et al.\ 1991), so the \xray\ emission in this region is probably associated both with the hot ejecta and the shocked ISM.

The long strip (``finger") that we detected in the soft band (Fig.~\ref{f:3bandsmpe}a) and in the median photon energy maps (Fig.~\ref{f:3bandsmpe}b) is likely an extension of the very dense small protrusion observed in the optical band by Danforth et al.\ (2001).
The densities of the cold component in regions 5\&6 (corresponding to this finger) are high, and this may also be related to the enhancement of 25 $\mu$m mid-infrared emission (Fig.~\ref{f:softvsIR}). However, no optical emission from the finger has been detected, thus suggesting that it does not correspond to the densest core of the XA region. The presence of the finger indicates that the protrusion in the cavity wall is much larger than what was thought before.

\subsection{The role of thermal conduction}
\label{The role of thermal conduction}

\begin{figure}
\centerline{ {\hfil\hfil \psfig{figure=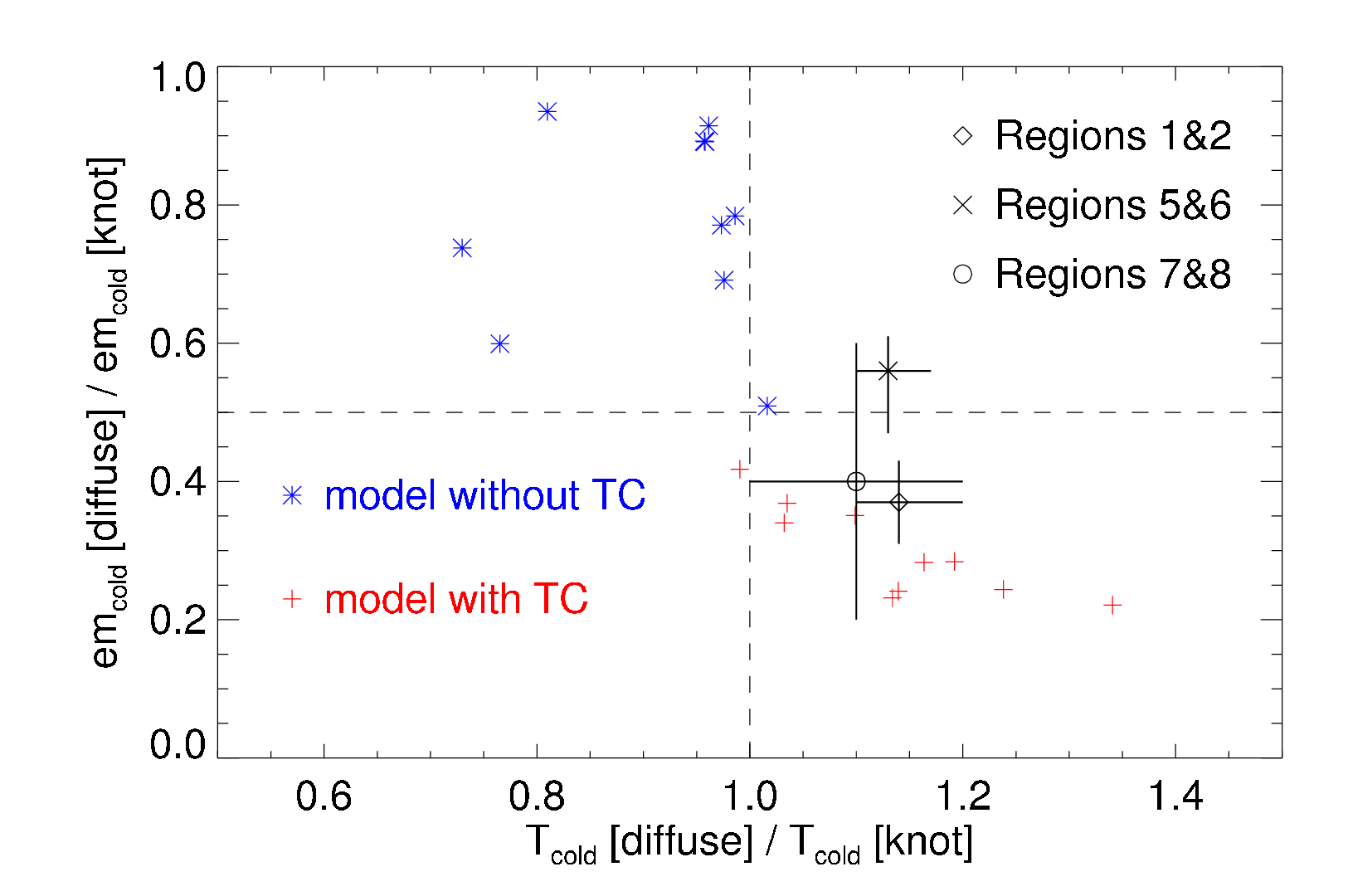,height=6.0cm,angle=0,clip=} \hfil\hfil}}
\caption{The figure is the same as Fig.~$7$ in Orlando et al.\ (2010),
but we added (in black) the points with the values derived from the analysis of the EPIC observation of the XA region in the \snr, by comparing the temperatures and the emission measures per unit area of the cold component derived for the knots and the corresponding diffuse regions. The blue stars are derived from the theoretical model without thermal conduction, the red crosses are derived from the model with thermal conduction.
} \label{f:TvsEM}
\end{figure}

To examine the role of thermal conduction in the shock-cloud interaction, we applied the method proposed by Orlando et al.\ (2010). By adopting this method, we can perform a direct and quick comparison between the observations and the results of the numerical model of shock-cloud interaction, without the need to run ad-hoc (and time-consuming) MHD simulations. In particular, we adopt a robust criterion that consists in comparing the temperature and the emission measure per unit area of the cold component derived from the knot (T$_{\rm cold}$[knot], em$_{\rm cold}$[knot]) to that derived from the corresponding diffuse regions (T$_{\rm cold}$[diffuse], em$_{\rm cold}$[diffuse]). Our results are shown in Fig.~\ref{f:TvsEM}. 
Orlando et al.\ (2010) showed that when thermal conduction is at work, we expect T$_{\rm cold}$[diffuse]/T$_{\rm cold}$[knot]$>1$ and em$_{\rm cold}$[diffuse]/em$_{\rm cold}$[knot]$<0.5$. As shown in Fig.~\ref{f:TvsEM}, we found that thermal conduction is efficient in all the shock-cloud interaction regions. Because of the presence of ejecta in regions 3\&4, we can not draw a conclusion there. In regions 5 and 6, that are at a relatively large distance from the border of the shell, we expect the emission to be contaminated by the contribution of multiple components along the LoS. This may be the reason of the high value em$_{\rm cold}$[diffuse]$/$em$_{\rm cold}$[knot] for these two regions. However, considering the error bars, all regions are consistent with a scenario where the contribution of thermal conduction is relevant.

Orlando et al.\ (2010) also showed that the slope of the $\overline{E}$ vs $R$ plot can provide important information on the efficiency of thermal conduction and on the physical properties of the shock and the cloud. We presented these plots for our spectral regions in Fig.~\ref{f:RvsE}. We found that the steepest descending trend of the ($\overline{E}$) vs $R$ plots (in regions 7\&8) is about two times flatter than that predicted by the theoretical model without thermal conduction (as shown in Fig.~$6$ in Orlando et al.\ 2010). Such flat descending trends are instead consistent with the shock-cloud interaction with prominent thermal conduction. As shown in Orlando et al.\ 2010, the $\overline{E}$ vs $R$ slope is sensitive to the shock Mach number, and the flatter trends in regions 1\&2 and 5\&6 probably indicate higher Mach numbers. In regions 7\&8, the steeper descending trend and the cold plateau provide convincing evidences for less efficient thermal conduction. This could be caused by slow shocks, presence of magnetic fields, or both. Considering that the magnetic field scale-length should be much larger than the distance between our regions, we do not expect large variations of the magnetic field and we conclude that a slow shock in regions 7\&8 may be more reasonably account for the steeper trend of the $\overline{E}$ vs $R$ plot.
The ascending trend in regions 3\&4 is not predicted in the theoretical models of shock-cloud interaction by Orlando et al. (2010). The ejecta origin of the X-ray emission in region 4 could be the reason of such ascending trend, but further investigation is needed to confirm this.

\section{Summary}

The \xmm\ observations toward the XA region in the \snr\ reveal several clumps interacting with the primary blast wave. Like other shock-cloud interaction regions, the spectra can be modeled with two thermal components, associated with the shocked cloud material and the cloud corona, respectively. We applied the diagnostic method proposed by Orlando et al.\ (2010) to investigate the physical process involved in the shock-cloud interaction. The primary results are:
\begin{itemize}
  \item[1.] the ISM around the XA region is dense and clumpy and the shocked cloud material occupies $\sim90\%$ of the volume behind the SNR shock front; the derived post-shock densities are in the range 3--11 cm$^{-3}$ for the low temperature component (associated with the cloud core) and 0.9--3.0 cm$^{-3}$ for the high temperature component (associated with the cloud corona).
\item[2.] the joint abundances of C, N, and O in the ISM are found to be lower than solar ($\sim0.2$ times the solar value), as in other parts of the \snr;
\item[3.] a long ``finger" ($\sim1.6$ pc) protruding from the cavity wall is detected;
\item[4.] thermal conduction is prominent in all the regions, and affects both the spatial distribution of the \xray\ emission and its spectral properties;
\item[5.] we detected the presence of ejecta and of a small ejecta shrapnel/protrusion.
\end{itemize}

\section*{Acknowledgments}
XZ would like to thank the friendly atmosphere at Osservatorio Astronomico di Palermo.
This work is partially supported by NSFC grants 10725312
and 10673003 and the China 973 Program grant 2009CB824800.

\label{lastpage}

\end{document}